\documentclass[12pt]{article}
\usepackage{amsmath,amssymb,amsopn,amsthm,times}

\title{Noncommutative Classical Dynamics on Velocity Phase Space and 
Souriau Formalism}

\author{
Jos\'e F. Cari\~nena$^*$, 
H\'ector Figueroa$^\dag $
and Partha Guha$^{\ddag\,\S}$
\\[1pc] 
$^*$\,Departamento de F\'{\i}sica Te\'orica, 
Universidad de Zaragoza,\\
50009 Zaragoza, Spain
\\[1pc]
$^\dag$\,Departamento de Matem\'aticas, Universidad de Costa Rica,\\
2060 San Pedro, Costa Rica  
\\[1pc]
$^\ddag$\, Institut des Hautes \'Etudes Scientifiques\\ Le Bois-Marie 35, 
route de Chartres 91440 Bures-sur-Yvette France \\
\\
$^\S$ S.N. Bose National Centre for Basic Sciences\\
JD Block, Sector-3, Salt Lake\\ Calcutta-700098, India }

\scrollmode
 
\advance\topmargin by -\headheight
\advance\topmargin by -\headsep
\evensidemargin=\oddsidemargin
\theoremstyle{plain}

\theoremstyle{definition}

\newcommand{\beqn}{\begin{equation}}
\newcommand{\eeqn}{\end{equation}}
\newcommand{\beqnarray}{\begin{eqnarray}}
\newcommand{\eeqnarray}{\end{eqnarray}}

\renewcommand{\a}{\alpha}          
\renewcommand{\b}{\beta}           
\DeclareMathOperator{\Div}{div}    
\newcommand{\dl}{\delta}           
\newcommand{\E}{\mathcal{E}}       
\newcommand{\eps}{\varepsilon}     
\newcommand{\F}{\mathcal{F}}       
\def\fd#1#2{\frac{d#1}{d#2}}       
\renewcommand{\H}{\mathcal{H}}     
\newcommand{\La}{\Lambda}          
\newcommand{\om}{\omega}         
\newcommand{\Om}{\Omega}           
\newcommand{\ox}{\otimes}          
\def\pd#1#2{\frac{\partial#1}{\partial#2}}
\newcommand{\R}{\mathbb{R}}        
\DeclareMathOperator{\Rot}{rot}    
\def\sd#1#2{\frac{d^2#1}{d#2^2}}   
\newcommand{\sepword}[1]{\qquad\hbox{#1}\qquad} 
\newcommand{\sg}{\sigma}           
\renewcommand{\th}{\theta}           
\newcommand{\thalf}{\tfrac{1}{2}}  
\newcommand{\V}{\mathcal{V}}       
\newcommand{\w}{\wedge}            
\newcommand{\wyw}{\wedge\cdots\wedge} 
\newcommand{\x}{\times}            
\newcommand{\Z}{\mathbb{Z}}        
\renewcommand{\:}{\colon}          
\def\<#1,#2>{\langle#1,#2\rangle}  

\makeatletter
\def\section{\@startsection{section}{1}{\z@}{-3.5ex plus -1ex minus
	 -.2ex}{2.3ex plus .2ex}{\large\bf}}
\def\subsection{\@startsection{subsection}{2}{\z@}{-3.25ex plus -1ex
	 minus -.2ex}{1.5ex plus .2ex}{\normalsize\bf}}
\makeatother

\InputIfFileExists{Whatnot.tex}{}{} 

\begin{document}

\maketitle

\begin{abstract}
We consider Feynman-Dyson's proof of Maxwell's equations using
the Jacobi identities on the velocity phase space. In this paper
we generalize the Feynman-Dyson's scheme by incorporating the non-commutativity
between various spatial coordinates along with the velocity
coordinates. This allows us to study a generalized class
of Hamiltonian systems.
We explore various dynamical flows associated to
the Souriau form associated to this generalized Feynman-Dyson's scheme.
Moreover, using the Souriau form we show that these new classes
of generalized systems are volume preserving
mechanical systems. 
\end{abstract}

\medskip


\medskip


\section{Introduction}

The study of exotic particle models with non-commutative position coordinates
was started in the last decade.
There are several physical phenomena appearing in condensed
matter physics, namely semiclassical Bloch electron phenomena,
fractional quantum Hall effect, double special relativity models, etc., 
that exhibit such feature.
All these models share the somewhat unusual feature that the 
Poisson brackets of the planar coordinates do not vanish. This class
of dynamical structures has appeared in geometric mechanics
and geometric control theory  \cite{Tulczyjew,Courant,MarsdenScheurle, Marsden}. 
In her thesis, S\'anchez de \'Alvarez \cite{SanchezdeAlvarez} indicates a 
characterization of the Poisson structure in terms of 
Poisson brackets of particular functions on the tangent bundle $TP$ of a Poisson manifold $P$, and discusses its functorial properties. 

A very noble derivation of a pair of Maxwell equations was originally
proposed by Feynman, but the exact details of his argument
came to the scientific community from the work of Freeman Dyson 
\cite{Dyson}. According to
Dyson, Feynman showed him the construction and examples of the
Lorentz force law and the homogeneous Maxwell equations in 1948. 
A derivation of a pair of Maxwell equations and  the
Lorentz force is based on the commutation relations between positions
and velocities for a single non-relativistic particle.
In general the locality property that different coordinates
commute is assumed. Due to increasing interest in non-commutative
field theories, it is worthwhile to consider the non-commutative
analogue of Feynman approach. This destroys the axiom of locality,
which, according to Dyson, was the original aim of Feynman.
Tanimura \cite{Tanimura} gave both a special relativistic and a general 
relativistic versions of Feynman's derivation. 
Land {\sl et al.} \cite{LSH} examined Tanimura's derivation in the framework of the proper time method in relativistic mechanics
and showed that Tanimura's result then corresponds to the five-dimensional electromagnetic theory previously derived 
from a Stueckelberg-type quantum theory in which one gauges the invariant parameter in the proper time method.
An extension of Tanimura's 
method has been achieved \cite{BGM99} by using the Hodge duality to derive 
the two groups of Maxwell's equations with a magnetic monopole in flat
and in curved spaces. A rigorous mathematical description of 
Feynman's derivation connected to the inverse problem for Poisson 
geometry has been formulated  in \cite{CIMS} (see also \cite{CIMM15}). Hughes \cite{Hughes} considered  
Feynman's derivation in the framework of the Helmholtz inverse
problem for the calculus of variations (see also \cite{MPL} and \cite{P09}).

In fact, it was pointed out by Jackiw that Heisenberg 
suggested in a letter to Peierls that spatial coordinates may
not commute, Peierls communicated the same idea to Pauli, who
informed it to Openheimer; eventually the idea arrived to
Snyder \cite{Snyder1,{Snyder2}} who wrote the first paper on the subject. 
Nowadays the physics in non-commutative planes is relevant not only in string 
theory but also in condensed matters physics \cite{SG04}.
In the context of the Feynman's derivation of electrodynamics,
it has been shown that non-commutativity
allows other particle dynamics than the standard formalism of
electrodynamics \cite{Felicia}. Noncommutative quantum mechanics is recently 
the subject of a wide range of works from particle physics to condensed matter 
physics. This has also been studied from the point of view of Feynman's 
formalism in \cite{LBMGG}.

The examples of exotic mechanics started to appear around 1995. Physicists 
obtained various models such that the Poisson brackets of the planar 
coordinates do not vanish. Souriau's orbit method 
\cite{DuvalHorvathy1, Horvathy,PH03} was used to construct a classical mechanics
associated with L\'evy-Leblond's exotic Galilean symmetry. In terms of the 
Souriau 2-form a wide set of Hamiltonian dynamical systems have been described 
in \cite{HMS,HMS1,Martina2,Martina4}. L\'evy-Leblond \cite{LevyLeblond} has realized that
due to the commutativity of the rotation group $O(2,\mathbb{R})$, the Lie 
algebra of the Galilei group in the plane admits a second exotic extension 
defined by 
$$
[K_1,K_2]=i\kappa\, I,
$$
where $\kappa$ is the new extension parameter. For a free particle
the usual equations of motions are unchanged and
$\kappa$ only contributes to the conserved quantities. It yields
the non-commutativity of the position coordinates.

Feynman procedure to obtain Maxwell's equation in electrodynamics
has been reviewed under different kind of settings, and several nontrivial 
and interesting generalizations are possible, see for instance
\cite{DuvalHorvathy,DuvalHorvathyb,Martina1,NairPolychronakos,Bracken,
Bracken2,BMG,KopfPaschke}.  Recently, Duval and Horv\'athy 
\cite{DuvalHorvathy} successfully applied the techniques of Souriau's 
orbit method \cite{Souriau,Souriau1,Souriau2} to various models. 
Incidentally, one of these models can be viewed as the non-relativistic 
counterpart of the relativistic anyon considered before by Jackiw and 
Nair \cite{NairPolychronakos}. Mathematically, the `exotic' model arises 
due to the particular properties of the plane. A wide set of dynamical 
systems can be derived from the Lagrange-Souriau $2$-form approach in 
three dimensions and the generalizations to higher number of degrees of 
freedom have been outlined in \cite{Martina2}.

Wong's equations describe the interaction
between the Yang-Mills field and an isotopic-spin carrying
particle in the classical limit. 
Feynman-Dyson's proof offers a way to check the consistency of 
these equations \cite{Lee}. See also \cite{BBGS} for a very recent paper.

In a slightly different context Kauffman \cite{Kauffman1}
introduced discrete physics based on a non-commutative 
calculus of finite differences. This gives a context for the Feynman--Dyson 
derivation of non-commutative electromagnetism. More recently, 
Kauffman~\cite{Kauffman2} found an interesting way to describe mechanics in 
a curved background interacting with gauge fields in such a way that the 
physical equations of motion emerge automatically from underlying algebraic 
relations in a non-commutative geometry and this construction depends largely 
on the Feynman-Dyson construction. In an interesting paper, Cortese and 
Garc\'{\i}a~\cite{CorteseGarcia} studied a variational principle for 
noncommutative dynamical systems in the configuration space. In particular 
they showed that the non-commutative consistency conditions (NCCC), that 
come from the analysis of the dynamical compatibility, are not the Helmholtz 
conditions of the generalized inverse problem of the calculus of variations. 
It has been shown that the $\theta$-deformed Helmholtz conditions are 
connected to a third-order time derivative system of differential equations.
Noncommutative phase spaces have been introduced by minimal couplings in 
\cite{Benin} and then  some of them are realized as coadjoint orbits of 
the anisotropic Newton-Hooke groups in two- and three-dimensional spaces.
This has been further generalized to realize noncommutative phase spaces as 
coadjoint orbit extensions of the Aristotle group in a two dimensional 
space \cite{Benin1}.

In this article we apply Souriau's orbit method to study
exotic mechanics on the tangent bundle or velocity
phase space. Souriau first unified both the
symplectic structure and the Hamiltonian into a single two-form.
It has an exotic symplectic form and a free Hamiltonian and
yields a generalized Hamiltonian mechanics. Duval and Horv\'athy used
Souriau's orbit method to construct a classical planar
system associated with L\'evy-Leblond's two-fold extended Galilean symmetry.
The four dimensional phase space is endowed with the following 
{\it exotic form}
$$
\Omega = d{p_i} \wedge d{q_i} + \frac{\theta}{2}\epsilon_{ij}\, dp_i \wedge dp_j,
$$
where summation on repeated indices is understood. The exotic term in the 
symplectic form only exists in the plane. 
Following~\cite{ZhouGuoWu,ZhouGuoPanWu} we also explore a volume-preserving 
flow on a symplectic manifold from the Souriau form associated with the 
velocity phase space.

Many authors \cite{DuvalHorvathy,DuvalHorvathyb,VSdS} have generalized this 
modification of the symplectic form by introducing the so-called dual 
magnetic field such that
$$ 
\Omega = d{p_i} \wedge d{q_i} + \frac{1}{2}g_{ij}\,dp_i \wedge dp_j 
+\frac{1}{2}f_{ij}\,dq_i \wedge dq_j.
$$
The coefficients  $g_{ij}$ and $f_{ij}$ are responsible of the noncommutativity
of momenta and positions, respectively. The classical dynamics in noncommutative 
space leads to noncommutative Newton's second law 
\cite{RomeroSantiagoVergara,WeiLong}. This generalization can be studied in
various types of noncommutative space-times; for instance Harikumar 
and Kapur studied in \cite{Harikumar} the modification to Newton's second 
law due to the kappa-deformation. In a very recent paper the modification 
of integrable models in the kappa-deformed scheme is analyzed 
\cite{GuhaHarikumar} and kappa-Minkowski space-time through exotic oscillator is studied in \cite{SG05}. 
Zhang {\sl et al.} \cite{ZhangHorvathy,ZH12} studied the 
$3D$ mechanics with non-commutativity, where the potential may also 
depend on the momentum. They obtained the conserved quantities by using 
van Holten's covariant framework. It is known that the Snyder model has 
the remarkable property of leaving the Lorentz invariance intact. 
Recently, motivated from loop quantum gravity an idea has been proposed to 
extend the Snyder model \cite{Snyder1} to space-times of constant curvature, 
by introducing a new fundamental constant whose inverse is proportional to 
the inverse of the cosmological constant. More recently, classical dynamics 
on Snyder space-times has drawn a lot of attention to physicists 
\cite{Banerjee,Ivetic,St12,LV}. Moreover Snyder dynamics in curved 
space-time has been extended
by Mignemi {\sl et al.} in \cite{MS14,M12}.  See also \cite{PGP14}.

\smallskip

The main theme of our paper is to show that non-commutativity
between coordinates allows us to construct various other
generalized classes of dynamical systems. Tools of non-commutative geometry 
often appear  in quantum gravity. Using a differential geometric theory on
non-commutative space-time Aschieri {\sl et al.}~\cite{Aschieri} defined 
$\theta$-deformed Einstein-Hilbert action, and by means of their technique 
of the deformation of the algebra of diffeomorphisms one can derive 
$\star$-deformed integrable systems~\cite{Guha} and Newtonian mechanics \cite{D12}.
Today we find non-commutativity in various fields of modern physics such as,
graphene, Hydrogen atom spectrum, etc. \cite{SMA,GT,ZD12}.

\smallskip

This paper is organized as follows: in order to the paper be self-contained, 
Section 2 is devoted to a review of multivector fields, Poisson bivector, 
Schouten--Nijenhuis bracket and various other geometrical tools. We give a 
brief geometrical description of Poisson manifolds in Section 3. Section 4 
is devoted to Souriau's formalism of generalized symplectic forms. We 
illustrate Souriau's construction through examples. Section 5 is focused on 
the construction of Feynman-Dyson's scheme and its connection to  Souriau's 
method. Section 6  relates volume preserving mechanical systems and Souriau's 
form. We finish our paper with an outlook in Section 7.

\section{Geometrical background}

Let $\F(M)$ be the algebra of $C^\infty$-class functions (the algebra of 
classical observables) on a manifold $M$ (the classical 
state space). We denote by $\Om^p(M)$ the space of $C^\infty$-class  
differentiable $p$-forms, and by $A^p(M)$ the space of $C^\infty$-class  
skew-symmetric contravariant tensor fields  of order $p$, often called 
$p$-vectors. By convention we set $A^0(M)=\Om^0(M)=\F(M)$ and 
$A^p(M)=\Om^p(M)=0$, when $p<0$. Then,
$$
\Om(M)=\bigoplus_{p\in\Z}\Om^p(M)\sepword{and}
A(M)=\bigoplus_{p\in\Z}A^p(M),
$$
are $\Z$-graded algebras under their exterior products; moreover both are
anticommutative, so, for instance, if $P\in A^p(M)$ and $Q\in A^q(M)$
$$
P\w Q=-(-1)^{pq}Q\w P.
$$
When $\a$ is a 1-form and $X$ is a vector field the $C^\infty(M)$-class function 
$\<\a,X>$ given by
$$
\<\a,X>(x):= \<\a(x),X(x)>:=\a(x)\bigl(X(x)\bigr),\quad \forall x\in M, 
$$
defines a pairing between $\Om^1(M)$ and $A^1(M)$. More generally
when $\eta$ in $\Om^q(M)$ and $P$ in $A^p(M)$ are decomposable, so
$\eta=\a_1\wyw\a_q$ and $P=X_1\wyw X_p$, for $\a_i$ in $\Om^1(M)$
and $X_j$ in $A^1(M)$, we set 
$$
\<\eta,P>:=\< \a_1\wyw\a_q,X_1\wyw X_p>=
\begin{cases} 0  & \text{if\;} p\ne q, \\
\det\bigl(\<\a_i,X_j>\bigr) & \text{if\;} p=q. 
\end{cases}
$$  
Since the value of $\<\eta,P>$ at a point only depends on the
value of $\eta$ and $P$ at this point, we can extend by bilinearity, 
in a unique
way, this pairing to arbitrary elements $\eta$ in $\Om(M)$ and
$P$ in $A(M)$. Furthermore, it is easy to check that if
$\eta$ is in $\Om^p(M)$, then 
$$
\<\eta,X_1\wyw X_p>=\eta(X_1,\dots,X_p).
$$
If $X$ is a vector field on $M$, the inner product $i(X)$ is a 
derivation of degree $-1$ on the graded algebra $\Om(M)$ and since the exterior
derivative $d$ is a derivation on $\Om(M)$ of degree 1, it
follows that the Lie derivative with respect to $X$, given by
Cartan's formula:
$$
\mathcal{L}_X:=[i(X),d]=i(X)\circ d+d\circ i(X),
$$
where $[\cdot,\cdot]$ means graded commutator, 
is a graded derivation  of degree 0 on $\Om(M)$. $\mathcal{L}_X$ can also be defined
on $A(M)$ as the unique derivation of degree 0 such that, for
$f$ in $A^0(M)$ and $Y$ in $A^1(M)$,
$$
\mathcal{L}_Xf=X(f) \sepword{and}\mathcal{L}_XY=[X,Y],
$$
where $[X,Y]$ is the usual Lie bracket on vector fields. Furthermore
the \textit{Schouten--Nijenhuis bracket} is defined as a natural 
extension of the Lie derivative with respect to a vector field 
on $A(M)$. More specifically,  it is defined as the unique bilinear
map $[\cdot,\cdot]_{SN}\:A(M)\x A(M)\to A(M)$ such that, for
$f$ and $g$ in $A^0(M)=\F(M)$, $X\in A^1(M)$, $P\in A^p(M)$, 
$Q\in A^q(M)$ and $R\in A^r(M)$,

\begin{description}

\item[a-)]\ $[f,g]_{SN}=0$

\item[b-)]\ $[X,Q]_{SN}=\mathcal{L}_XQ$ 

\item[c-)]\ $[P,Q]_{SN}=-(-1)^{(p-1)(q-1)}[Q,P]_{SN}$ 

\item[d-)]\ $[P,Q\w R]_{SN}=[P,Q]_{SN}\w R+(-1)^{(p-1)q}Q\w[P,R]_{SN}$

\end{description}

From these properties it readily follows that $[P,Q]_{SN}$ belongs to
$A^{p+q-1}(M)$, therefore the last property means that the endomorphism
$d_P\:A(M)\to A(M)$ given by
\begin{equation}
d_PQ:=[P,Q]_{SN},
\label{eq:derivation}
\end{equation}
is a derivation of $A(M)$ of degree $p-1$. A somewhat long, but otherwise
easy, induction, based on the defining properties, gives
\begin{align} 
(-1)^{(p-1)(r-1)}\bigl[P,[Q,R]_{SN}\bigr]_{SN}
&+(-1)^{(q-1)(p-1)}\bigl[Q,[R,P]_{SN}\bigr]_{SN} \notag \\
&+(-1)^{(r-1)(q-1)}\bigl[R,[P,Q]_{SN}\bigr]_{SN}=0,
\label{eq:gjacobi}
\end{align}
which is called graded Jacobi identity.

This, together with bilinearity, \textbf{c-)}, and the fact that 
$[P,Q]_{SN}$ belongs to $A^{p+q-1}(M)$ means that $A(M)$, equipped with 
the Schouten--Nijenhuis bracket, is a graded Lie algebra when the degree 
of $P$ in $A^p(M)$ is declared to be $p-1$, not $p$. So, for instance, 
vector fields would be the homogeneous elements of degree 0 under this 
new grading of $A(M)$.
To perform computations with the Schouten--Nijenhuis bracket it is 
convenient to extend the definition of the interior product. If $\eta$ 
is in $\Om(M)$, $f$ is a function and $X_1,\dots,X_p$ are vector fields 
we set 
$$
i(f)\eta:= f\eta\sepword{and}i(X_1\wyw X_p)
\eta:= 
i(X_1)\circ\cdots\circ i(X_p)\eta.
$$
In general, $i(P)$ is defined in such a way that the map 
$i(\cdot)\:A(M)\to\E\bigl(\Om(M)\bigr)$, where $\E\bigl(\Om(M)\bigr)$ is 
the space of endomorphism of $\Om(M)$, is $\F(M)$-linear. In particular, 
$i(P\w Q)\eta=i(P)(i(Q)\eta)$, for all $P$ and $Q$ in $A(M)$. Furthermore, 
when $\eta$ is a $p$-form
$$
i(X_1\wyw X_p)\eta= i(X_1)\circ\cdots\circ i(X_p)\eta=\eta(X_p,\dots,X_1)
=(-1)^{\frac{(p-1)p}{2}}\eta(X_1,\dots,X_p),
$$
therefore for any $P\in A^p(M)$
\begin{equation}
i(P)\eta=(-1)^{\frac{(p-1)p}{2}}\<\eta,P>.
\label{eq:intproduct}
\end{equation}
Unfortunately $i(P)$, in general, is not a derivation, which complicates
computations. Nevertheless, another simple induction gives
\begin{equation}
i\bigl([P,Q]_{SN}\bigr)=\bigl[[i(P),d],i(Q)\bigr],
\label{eq:tobeexpected}
\end{equation}
where the brackets on the right are the usual brackets on the
algebra of endomorphisms of $A(M)$. Notice that when $P=X$ is a vector 
field this reduces to the well-known relation among interior
products and Lie derivatives: $i(\mathcal{L}_XQ)=[\mathcal{L}_X,i(Q)]$.

If $P$ is a $p$-vector and $\eta$ is a $(p-1)$-form, then $i(P)\eta=0$
and $i(P)\circ i(f)\eta=i(P)(f\eta)=0$. These, together 
with~\eqref{eq:intproduct} and~\eqref{eq:tobeexpected} entail
\begin{align*}
\<\eta,[P,f]_{SN}> &=(-1)^{\frac{(p-2)(p-1)}{2}}i([P,f]_{SN})\eta \\
&=(-1)^{\frac{(p-2)(p-1)}{2}}\bigl[[i(P),d],i(Q)\bigr]\eta  \\
&=(-1)^{\frac{(p-2)(p-1)}{2}}\Bigl(i(P)\circ d\circ i(f)\eta
-(-1)^pd\circ i(P)\circ i(f)\eta  \\
&\hskip 3cm-i(f)\circ i(P)\circ d\eta
-(-1)^pi(f)\circ d\circ i(P)\eta\Bigr)  \\ 
&=(-1)^{\frac{(p-2)(p-1)}{2}}\Bigl(i(P)\circ d\circ i(f)\eta
-i(f)\circ i(P)\circ d\eta\Bigr)  \\
&=(-1)^{\frac{(p-2)(p-1)}{2}}i(P)(df\w\eta) \\
&=(-1)^{(p-1)^2}\<df\w\eta,P>  \\
&=(-1)^{(p-1)(p-2)}\<\eta\w df,P>  \\
&=\<\eta\w df,P>.
\end{align*}
Repited use of this gives
\begin{align}
\<df_1\wyw df_p,P>&=\<df_1\wyw df_{p-1},[P,f_p]_{SN}> \notag\\
&=\cdots=\Bigl[\cdots\bigl[[P,f_p]_{SN},f_{p-1}\bigr]_{SN},\cdots,f_1\Bigr]_{SN}.
\label{eq:nicefor}
\end{align}

\section{Poisson Manifolds}

A \textit{Poisson structure} on $M$ is a 
 skew-symmetric $\mathbb{R}$-bilinear map 
$\{\cdot,\cdot\}\:\F(M)\x\F(M)\to\F(M)$
satisfying the Jacobi identity:
$$
\{f,\{g,h\}\}+\{h,\{f,g\}\}+\{g,\{h,f\}\}=0,\quad
\forall f,g,h\in \F(M)\,,
$$ 
and such that the map $X_f =\{\cdot,f\}$ is a
derivation of the associative and commutative  algebra $\F(M)$, for each 
$f\in \F(M)$, or in other words, $X_f$ is a vector field, 
usually called a \textit{Hamiltonian vector field}, and 
$f$ is said to be the \textit{Hamiltonian} of $X_f$. 
This property characterizing derivations of the associative and commutative  
algebra $\F(M)$, $\{g_1g_2,f\}=g_1\{g_2,f\}+g_2\{g_1,f\}$, called  
Leibniz' rule, is very important and gives a compatibility condition of the
associative and commutative  algebra structure in $\F(M)$ with the Lie algebra 
given in  $\F(M)$ by the Poisson bracket.

To construct Poisson structures let $\La$ be an element 
of $A^2(M)$, if $f\in A^0(M)$ and $g\in A^0(M)$ are two 
functions, using~\eqref{eq:nicefor}, we define a third 
function by
\begin{align}
\{f,g\}:=\La(df,dg)=-\La(dg,df)=-\<dg\w df,\La>
=-\bigl[[\La,f]_{SN},g\bigr]_{SN}.
\label{eq:Pbracket}
\end{align}
By construction $X_f:=[\La,f]_{SN}$ is a vector field, 
and the defining property \textbf{b-)} entails
\begin{align}
X_f(g)=\mathcal{L}_{X_f}g=[X_f,g]_{SN}=-\{f,g\}=\{g,f\}.
\label{eq:Pbracketagain}
\end{align}
In particular
$$
\{g,\{h,f\}\}=\Bigl[[\La,g]_{SN},\bigl[[\La,h]_{SN},f\bigr]_{SN}\Bigr]_{SN}
=\bigl[X_g,[X_h,f]_{SN}\bigr]_{SN}=\mathcal{L}_{X_g}\circ\mathcal{L}_{X_h}f.
$$
By the same token
$\{h,\{f,g\}\}=-\{h,\{g,f\}\}=-\mathcal{L}_{X_h}\circ\mathcal{L}_{X_g}f$. On the other hand,
the graded Jacobi identity~\eqref{eq:gjacobi}, the defining property 
\textbf{b-)},  and~\eqref{eq:Pbracketagain} 
entail
\begin{align*}
\{f,\{g,h\}\}&=-\{\{g,h\},f\}=\{[X_g,h]_{SN},f\} 
=-\Bigl[\bigl[\La,[X_g,h]_{SN}\bigr]_{SN},f\Bigr]_{SN} \\
&=-\Bigl[\bigl[X_g,[h,\La]_{SN}\bigr]_{SN},f\Bigr]_{SN}
-\Bigl[\bigl[h,[\La,X_g]_{SN}\bigr]_{SN},f\Bigr]_{SN}  \\
&=-\bigl[[X_g,X_h]_{SN},f\bigr]_{SN}
-\Bigl[\bigl[h,[\La,X_g]_{SN}\bigr]_{SN},f\Bigr]_{SN}  \\
&=-\mathcal{L}_{[X_g,X_h]}f
-\Bigl[\bigl[h,[\La,X_g]_{SN}\bigr]_{SN},f\Bigr]_{SN}. 
\end{align*}
Altogether gives
\begin{align*}
\{f,\{g,h\}\}+\{g,\{h,f\}\}+\{h,\{f,g\}\}
&=\Bigl(\mathcal{L}_{X_g}\circ\mathcal{L}_{X_h}-\mathcal{L}_{X_h}\circ\mathcal{L}_{X_g}-\mathcal{L}_{[X_g,X_h]}\Bigr)f\\
&\quad-\Bigl[\bigl[h,[\La,X_g]_{SN}\bigr]_{SN},f\Bigr]_{SN}  \\
&=-\Bigl[\bigl[h,[\La,X_g]_{SN}\bigr]_{SN},f\Bigr]_{SN}.
\end{align*}
Furthermore, from the graded Jacobi identity
\begin{align}
0&=\bigl[\La,[\La,g]_{SN}\bigr]_{SN}+\bigl[\La,[\La,g]_{SN}\bigr]_{SN}
+\bigl[g,[\La,\La]_{SN}\bigr]_{SN}  \notag\\
&=2\bigl[\La,[\La,g]_{SN}\bigr]_{SN}+\bigl[g,[\La,\La]_{SN}\bigr]_{SN} 
\notag\\
&=2[\La,X_g]_{SN}+\bigl[g,[\La,\La]_{SN}\bigr]_{SN},
\label{eq:thetrick}
\end{align}
therefore, by~\eqref{eq:nicefor}
\begin{align}
\{f,\{g,h\}\}+\{g,\{h,f\}\}+\{h,\{f,g\}\}
&=\frac{1}{2}
\biggl[\Bigl[h,\bigl[g,[\La,\La]_{SN}\bigr]_{SN}\Bigr]_{SN},f\biggr]_{SN}  
\notag \\
&=\frac{1}{2}
\biggl[\Bigl[h,\bigl[[\La,\La]_{SN},g\bigr]_{SN}\Bigr]_{SN},f\biggr]_{SN} 
\notag \\
&=-\frac{1}{2}
\biggl[\Bigl[\bigl[[\La,\La]_{SN},g\bigr]_{SN},h\Bigr]_{SN},f\biggr]_{SN}  
\notag \\
&=-\frac{1}{2}\<df\w dh\w dg,[\La,\La]_{SN}>.
\label{eq:wonderful}
\end{align}
It follows that the bracket defined via a bivector field $\La$ is a Poisson 
structure exactly when $[\La,\La]_{SN}=0$, and when this happens we say 
that $\La$ is a \textit{Poisson tensor}. This elementary, but clever, 
computation was first performed by Lichnerowicz in~\cite{Lichnerowicz}, 
who also realized that, when $\La$ is a Poisson tensor and $P$ is in 
$A^p(M)$, the graded Jacobi identity implies
\begin{align*}
0&=(-1)^{p-1}\bigl[\La,[\La,P]_{SN}\bigr]_{SN}
-\bigl[\La,[P,\La]_{SN}\bigr]_{SN}
+(-1)^{p-1}\bigl[P,[\La,\La]_{SN}\bigr]_{SN}  \\
&=2(-1)^{p-1}\bigl[\La,[\La,P]_{SN}\bigr]_{SN}.
\end{align*}
In other words, for Poisson tensors, the derivation $d_\La$, defined as 
in~\eqref{eq:derivation} by $d_\La P=[\La,P]_{SN}$, satisfies the cocycle 
condition
$$
d_\La\circ d_\La=0.
$$
On the other hand, from~\eqref{eq:thetrick} we see that, for
Poisson tensors, $[X_f,\La]_{SN}=0$, this together with the
graded Jacobi identity and~\eqref{eq:Pbracketagain} give
\begin{align}
[X_f,X_g]=\bigl[X_f,[\La,g]_{SN}\bigr]_{SN}
&=-\bigl[\La,[g,X_f]_{SN}\bigr]_{SN}+\bigl[g,[X_f,\La]_{SN}\bigr]_{SN}  
\notag\\
&=\bigl[\La,[X_f,g]_{SN}\bigr]_{SN}=-[\La,\{f,g\}]   \notag\\ 
&= X_{-\{f,g\}}.
\label{eq:Liemorphism}
\end{align} 
 
The converse is also true: a Poisson structure determines a
Poisson tensor. To see this let $\xi_a$ denote a set of local 
coordinates on $M$, then, using the summation index convention,
$$
X_f =X_f(\xi_a)\pd{}{\xi_a}=\{\xi_a,f\}\pd{}{\xi_a},
$$
hence 
$$
\{f,g\}=X_g(f)=\{\xi_a,g\}\pd{f}{\xi_a}\,.
$$ 
Thus,
\begin{equation}
\{\xi_a,g\}=-\{g,\xi_a\}=-\{\xi_b,\xi_a\}\pd{g}{\xi_b}
=\{\xi_a,\xi_b\}\pd{g}{\xi_b},
\label{eq:localformula}
\end{equation}
and the local coordinate expression of the Poisson Bracket becomes
\begin{equation}
\{f,g\}=\{\xi_a,\xi_b\}\pd{g}{\xi_b}\pd{f}{\xi_a}.
\label{eq:Poissonbracket}
\end{equation}
Therefore to compute the Poisson bracket of any pair of
functions is enough to know the fundamental Poisson 
brackets
$$
\Lambda_{ab} = \{\xi_a,\xi_b\}.
$$
Moreover, the value of $\{f,g\}$ at a point $m\in M$
does not depend neither on $f$ nor on $g$ but only on  $df$ and $dg$, as 
explicitly shown in~\eqref{eq:Poissonbracket}, hence from the
Poisson structure we get a twice contravariant skew-symmetric tensor 
$$
\Lambda(df,dg) := \{f,g\}.
$$
Indeed, the local coordinate expression of $\La$ is
$$
\La= \La_{ab}\pd{}{\xi_a}\w\pd{}{\xi_b},
$$
and if $\bar\xi=\phi(\xi)$ is another set of
local coordinates on $M$, then,
$$
\bar \Lambda_{ab}= \{\bar\xi_a,\bar\xi_b\}
= \{\phi_a,\phi_b\}
= \{\xi_c,\xi_d\}\pd{\phi_a}{\xi_c}\pd{\phi_b}{\xi_d}
= \Lambda_{cd}\pd{\phi_a}{\xi_c}\pd{\phi_b}{\xi_d}\,,
$$
so the components of $\Lambda$ do change like the local
coordinates of a skew-symmetric twice contravariant tensor
which, by~\eqref{eq:wonderful}, it is a Poisson tensor. We are 
using the convention that in the local expression of the wedge 
product only summands whose subindex on the left hand side term 
is smaller than the subindex on the right hand side term do appear.

For any function $h\in \F(M)$ the integral curves of the  dynamical vector 
field $X_h$ are precisely determined by the solutions of the 
system of differential equations
\begin{equation}
\fd{\xi_a}{t} =\{{\xi_a},h\}\,,
\label{eq:timeevolution}
\end{equation}
and the dynamical evolution of a function $f$ on $M$ is given by
$$
\fd{f}{t} = \{f,h\},
$$
or in local coordinates
$$
\fd f{t} = \Lambda_{ab}\pd f{\xi_a}\pd{h}{\xi_b}.
$$

Interesting examples of Poisson manifolds are symplectic
manifolds. A symplectic form $\om$ on $M$ determines a bundle map 
$\om^{\flat}\:TM\to T^*M$ over the identity, which gives rise to the 
corresponding linear map between their spaces of sections, defined by
$$
\bigl(\om^{\flat}(X)\bigr)(Y):=\<\om^{\flat}(X),Y>=\om(X,Y).
$$
Since $\om$ is non-degenerate $\om^{\flat}$ is actually
an isomorphism; we denote $\om^\sharp$ the inverse map and
 define a bivector $\La$ by
$$
\La(\a,\b):=\om\bigl(\om^\sharp(\a),\om^\sharp(\b)\bigr),
$$
if $\a$ and $\b$ are 1-forms. When $\a=df$ is exact the
corresponding vector field is denoted by $X_f:=\om^\sharp(df)$,
and we say $X_f$ is the \textit{vector field associated to  $f$ with 
respect to $\om$}. It is actually defined by the 
equation $i(X_f)\om=df$. Furthermore, let $\{\cdot,\cdot\}$ be the 
bracket associated to $\La$ via~\eqref{eq:Pbracket}. Then the 
vector field associated to $f$ with respect to $\om$ is 
also the Hamiltonian vector field given in~\eqref{eq:Pbracketagain}, 
explaining why we use the same notation. If $\xi_a$, $a=1, \ldots ,m$, 
denote local coordinates and 
$\partial/\partial\xi_1,\dots,\partial/\partial\xi_m$,
and $d\xi_1,\dots,d\xi_m$ are, respectively, the local basis of $A^1(M)$
and $\Om^1(M)$ associated to $\xi_a$, let $B=(B_{ab})$ be the matrix of 
the linear map $\om^{\flat}$ relative to these bases, and 
$\om=\om_{ab}\,d\xi_a\w d\xi_b$ the local expression of the 
symplectic form, then
$$
\om_{ab}=\om\left(\pd{}{\xi_a},\pd{}{\xi_b}\right)
=\biggl(\om^{\flat}\left(\pd{}{\xi_a}\right)\biggr)\left(\pd{}{\xi_b}\right)
=\Bigl(B_{ac}d\xi_c\Bigr)\left(\pd{}{\xi_b}\right)=B_{ab}.
$$
Thus, $B=(\om_{ab})$, and the matrix of $\om^\sharp$ associated to
these bases is the inverse of $B$. On the other hand,
\begin{align*}
d\om(X_f,X_g,X_h)&= X_f\bigl(\om(X_g,X_h)\bigr)
+X_g\bigl(\om(X_h,X_f)\bigr)+X_h\bigl(\om(X_f,X_g)\bigr)  \\
&\quad-\om([X_f,X_g],X_h)-\om([X_g,X_h],X_f)-\om([X_h,X_f],X_g).
\end{align*}
Nevertheless, $\om(X_g,X_h)= \La(dg,dh)=\{g,h\}$,
so~\eqref{eq:Pbracketagain} entails
$$
X_f\bigl(\om(X_g,X_h)\bigr)=X_f(\{g,h\})=-\{f,\{g,h\}\}.
$$
Also, by~\eqref{eq:Liemorphism},
$[X_g,X_h]=[X_g,X_h]_{SN}=X_{-\{g,h\}}$, hence
$$
\om([X_g,X_h],X_f)=\om(X_{-\{g,h\}},X_f)=-\{\{g,h\},f\}
=\{f,\{g,h\}\}.
$$
It follows that
\begin{equation}
0=d\om(X_f,X_g,X_h)=-2(\{f,\{g,h\}\}+\{g,\{h,f\}\}+\{h,\{f,g\}\}),
\label{hownice}
\end{equation}
so $\La$ is a Poisson tensor. 

Reciprocally, from a Poisson tensor
$\La$ we get a bundle map $\La^{\sharp}\:T^*M\to TM$, defined by
$$
\<\a,\La^{\sharp}(\b)>:=\La(\a,\b).
$$
In general $\La^{\sharp}$ is not a bundle isomorphism. We say the
Poisson structure is \textit{regular} when that is the case, and
then we denote the inverse map by $\La^{\flat}$. Thus we have an 
identification of $T_xM$ and $T_x^*M$, for each point $x$ of $M$
and, therefore, an identification of higher order contravariant 
and covariant tensors. In particular, we define a 2-form $\om$ by
$$
\om(X,Y):=\La\bigl(\La^{\flat}(X),\La^{\flat}(Y)\bigl).
$$
Notice that, from this point of view, $X_f=\La^{\sharp}(df)$.
Indeed, from~\eqref{eq:Pbracketagain}
$$
\<dg,X_f>= X_f(g)=\{g,f\}=\La(dg,df)=\<dg,\La^{\sharp}(df)>.
$$
Therefore,
$$
\om(X_f,X_g)= \La\Bigl(\La^{\flat}\bigl(\La^{\sharp}(df)\bigr),
\La^{\flat}\bigl(\La^{\sharp}(dg)\bigr)\Bigr)=\La(df,dg).
$$
In particular the brackets associated to $\om$ and $\La$ coincide.
Since the Poisson bracket satisfies the Jacobi identity, 
~\eqref{hownice} entails, $d\om(X_f,X_g,X_h)=0$. Given that 
locally one can consider a basis consisting of Hamiltonian vector 
fields, we conclude that $\om$ is a closed 2-form, moreover by definition
it is non-degenerate, hence $\om$, so defined, is indeed a symplectic form.
Using local coordinates as above, we see that the matrix of $\La^{\sharp}$
with respect to the standard bases is $(\La_{ab})$, where 
$\La= \La_{ab}\pd{}{\xi_a}\w\pd{}{\xi_b}$
is the local expression of $\La$.

\section{Souriau's prescription}

As far as we know, Souriau~\cite{Souriau,Souriau1,Souriau2} was the first 
to realize that since a dynamical system has two pieces, the symplectic 
form and the Hamiltonian, the equations of motion can be described by 
different data, modifying one or the other component. Thus, a perturbed 
dynamical system can be described starting from the free case by modifying 
the Hamiltonian, as was classically done, or simply by changing the 
symplectic form (see also \cite{GS84}). This idea of
adding an extra term to the symplectic form was successfully exploited 
by Souriau in his study of the orbit method, and it is what is behind
the exotic mechanics, and several other models where non-commutativity
of the variables is employed. But before we tackle that, let us consider 
a more down to earth example.

The classical method to derive the Lorentz equations in a relativistically 
invariant way is to use the so called \textit{minimal coupling}, which
consists in making the substitution $p\mapsto p-eA$ inside the free
Hamiltonian, where $A$ is the vector potential of the electromagnetic field
and $e$ is the electric charge. Thus, the starting point is the 
cotangent bundle $T^*M$, of a manifold $M$, endowed with its canonical 
symplectic form $\om_0=d\th_0$, where $\th_0$ is the canonical 1-form given,
in local cotangent bundle coordinates $(q_i,p_i)$, induced from local 
coordinates $(q_i)$ on $M$,  by $\th_0=p_i\,dq_i$, together with a Hamiltonian 
$H\:T^*M\to\mathbb{R} $, which one replaces by $H_A=H\circ\phi_A^{-1}$,
where $\phi_A\:T^*M\to T^*M$ is the bundle map over the identity given by 
$\phi_A(q,p)=\bigr(q,p+eA(q)\bigl)$, and $A=A_i(q)\,dq_i$ is a basic 1-form  
on $T^*M$. The Hamiltonian vector field $X_{H_A}$ associated to this new 
Hamiltonian $H_A=(\phi_A ^{-1} )^*H$, that leads to the equation of motion, 
is given by
$$
i(X_{H_A})d\th_0=dH_A=(\phi_A ^{-1} )^*(dH).
$$
On the other hand, and with an abuse of notation, we denote  $\sg$ both the
1-form on $M$ defined by $\sg=eA^i(q)\,dq_i$, as well as the basic 1-form on 
$T^*M$ obtained by pulling back $\sg$ by the canonical projection 
$\pi\:T^*M\to M$. Then,
$$
\phi_A^*(d\th_0)=\phi_A^*(dp_i\w dq_i)= d\phi_A^*(p_i)\w d\phi_A^*(q_i)
=d\bigr(p_i+eA_i(q)\bigl)\w\, dq_i=d\th_0+d\sg,
$$
and since $\phi$ is a diffeomorphism,
$$
i\bigl(\phi_{A*}(X_{H_A})\bigr)(d\th_0+d\sg)
=i\bigl(\phi_{A*}(X_{H_A})\bigr)(\phi_A^*d\th_0)
=\phi_A^*(i(X_{H_A})d\th_0)=\phi_A^*\bigl((\phi_A^*)^{-1}(dH)\bigr)
=dH.
$$
In other words, by adding the extra term $d\sg$ to the symplectic
form, which, by the way, it is a basic 2-form (i.e. locally it is
of the form $g_{ij}(q)\,dq_i\w dq_j$, so it does not involve the $p$'s), 
we see that the vector field $\phi_{A*}(X_{H_A})$ is the Hamiltonian 
vector field associated to the original Hamiltonian $H$ with respect 
to this new symplectic form, and we obtain the same equations of 
motion.

 If $\om=\om_0+\thalf g_{ij}\,dq_i\w dq_j$, where $g_{ij}(q,p)$ is
a skew-symmetric matrix, and the Hamiltonian vector field $X_H$ is  
$X_H=V_i\partial_{q_i}+W_i\partial_{p_i}$, then
\begin{align*}
i(X_H)\om&=(i(X_H)dq_i) \w dp_i- dq_i\w( i(X_H)dp_i)
+\frac{1}{2}g_{ij}(i(X_H)dq_i)\w dq_j\\
&\qquad-\frac{1}{2}g_{ij}\,dq_i\w (i(X_H)dq_j)\\  
&=-W_i\,dq_i+V_i\,dp_i+g_{ij}\,V_j\,dq_i
\end{align*}
The equation $i(X_H)\om=dH$, entails
$$
V_i=\pd{H}{p_i}\sepword{and}W_i=-\pd{H}{q_i}+g_{ki}\pd{H}{p_k},
$$
therefore by~\eqref{eq:Pbracketagain} the Poisson bracket associated 
to $\om$ is given by
$$
\{F,H\}=\pd{H}{p_i}\pd{F}{q_i}-\pd{H}{q_i}\pd{F}{p_i}
+g_{ij}\pd{H}{p_i}\pd{F}{p_j}
=\{F,H\}_0+g_{ij}\pd{H}{p_i}\pd{F}{p_j} \equiv {\mathbb X}_H(F),
$$
where $\{\cdot,\cdot\}_0$ stands for the Poisson bracket corresponding to 
$\om_0$. It follows that the generalized (Hamiltonian) vector field is
$$
{\mathbb X}_{H} = X_H + g_{ij}\pd{H}{p_i}\pd{}{p_i}.
$$
The equations of motion are given by
$$
\fd{q_k}{t} = \pd{H}{p_k}, \qquad \fd{p_k}{t} 
= -\pd{H}{q_k} + g_{ik}\pd{\H}{p_i}.
$$
Our construction can be extended easily to a more general Souriau form 
$$ 
\omega = \omega_0 + \thalf g_{ij}\,dq_i\w dq_j 
+ \thalf f_{ij}\,dp_i\w dp_j,
$$
and the equations of motion are then given by
$$
\fd{q_k}{t} = \pd{H}{p_k} + f_{ki}\,\pd{H}{q_i}, 
\qquad \fd{p_k}{t} = -\pd{H}{q_k} + g_{ik}\,\pd{H}{p_i}.
$$
Then if we assume the Hamiltonian is of the form 
$$
H (q,p)= \frac{\delta^{ij}p_ip_j}{2m} + \V(q),
$$
with the potential energy $\V$ depending only on the configuration coordinates
$q_i$, the equations of motion are 
$$
\fd{q_k}{t} = \frac{p_k}{m} + f_{ki}\,\pd{\V}{q_i}, \qquad 
\fd{p_k}{t} = -\pd{\V}{q_k} + g_{ik}\,\frac{p_i}{m}.
$$
These are equivalent to the modified Newton's second law 
\cite{DuvalHorvathyb,Benin,RomeroSantiagoVergara,WeiLong}
\begin{equation}\label{Newtonseclaw}
m\sd{q_k}{t} = - \pd{\V}{q_k} + g_{ik}\frac{p_i}{m} + 
m\fd{}{t} \left(f_{ki}\pd{\V}{q_i} \right).
\end{equation}
The second term of equation (\ref{Newtonseclaw}) is a correction due to the 
noncommutativity of momenta and the third term is that of noncommutativity of 
coordinates. It follows that even for the case $\mathcal{V} = 0$ the 
particle accelerates because of the noncommutativity of momenta.

The second procedure has the advantage that it works even 
when only the 2-form  is globally defined, with  no reference to the 1-form
$\th_0$ made. In~\cite{Sternberg} this idea was generalized to the 
case of a classical particle in the presence of a Yang-Mills field.
When the Poisson manifold $M$ is the tangent bundle $TQ$ of a 
$n$-dimensional manifold, so $M$ is known as the \textit{velocity phase 
space}, Souriau also proposed to describe the dynamics not on phase space 
but in what he called \textit{evolution space}, with coordinates 
$(x_i,\dot x_j,t)$. His idea was to join  the symplectic form 
$\om$ on phase space with the Hamiltonian by considering the two-form 
$\om - dH\w dt$ on the evolution space, and then perform the minimal 
coupling recipe. This allows to recover the Euler-Lagrange equations, 
and it is equivalent to Faddeev-Jackiw construction 
\cite{DuvalHorvathy,FaddeevJackiw}. Recently, Bolsinov and 
Jovanovi\'c~\cite{BolsinovJovanovic} considered $G$-invariant magnetic 
geodesic flows on coadjoint orbits of a compact Lie group $G$, where 
$\sg$ is the Kirillov-Kostant two-form.

\subsection{Souriau's formalism and exotic mechanics}

For concreteness let us ponder the `exotic' plane studied by Horv\'athy
in~\cite{Horvathy}. Thus, we consider the  dynamical system 
$(T^*\R^2,\om_\vartheta,H_0)$, where $\vartheta\in\mathbb{R}$,  
$$
\om_\vartheta=dq_1\w dp_1+dq_2\w dp_2-\vartheta\, dp_1\w dp_2\sepword{and} 
H_0=\frac{p_1^2+p_2^2}{2m}.
$$
The 2-form $\om_\vartheta$ is not only closed but exact, and as the 
associated map $\om^\flat_\vartheta\:A^1(T^*\R^2)\to\Om^1(T^*\R^2)$ is given 
by the matrix 
$$
\om^\flat_\vartheta=
\begin{pmatrix}
0&0&-1&\hskip 3mm 0  \\  0&0&\hskip 3mm 0&-1 \\
1&0&\hskip 3mm 0&\hskip 3mm \vartheta\\ 0&1&-\vartheta&\hskip 3mm 0
\end{pmatrix},
$$
which is regular for any value of $\vartheta$,  the 2-form  $\om_\vartheta$
is symplectic.
The inverse matrix is 
$$
\om^\sharp_\vartheta=
\begin{pmatrix}\hskip 3mm 0&\hskip 3mm \vartheta&1&0 \\
- \vartheta&\hskip 3mm 0&0&1\\ -1&\hskip 3mm0&0&0\\ 
\hskip 3mm 0&-1&0&0
\end{pmatrix},
$$
which corresponds to the Poisson structure associated to the bi-vector
$$
\Lambda_\vartheta= \vartheta \pd{}{q_1}\wedge
\pd{}{q_2}+\pd{}{q_1}\wedge \pd{}{p_1}
+\pd{}{q_2}\wedge \pd{}{p_2}\,,
$$
so the fundamental Poisson commutators are
$$
\{q_1,q_2\}=\vartheta\,,\quad \{q_1,p_1\}=\{q_2,p_2\}=1\,,\quad
\{q_1,p_2\}=\{q_2,p_1\}=0 \,,\quad \{p_1,p_2\}=0\,,
$$
and the dynamical vector field is given by
$$
\dot q_1=\{q_1,H_0\}=\frac{p_1}m,\quad \dot q_2=\{q_2,H_0\}=\frac{p_2}m,\quad
\dot p_1=\{p_1,H_0\}=0,\quad \dot p_2=\{p_2,H_0\}=0\,.
$$
In this way we obtain a 1-parameter family of symplectic structures
for the free particle.

These symplectic structures are the sum of two symplectic structures
that are invariant under rotations in the plane, one of them being the
canonical symplectic structure on $T^*\R^2$. The generating function  
of such 1-parameter group will be the function $f$ satisfying 
$$
i(X)\om_\vartheta=df,
$$
where $X$ is the vector field that is the cotangent lift of the rotation 
generator in configuration space, 
$$
X=q_1\pd{}{q_2}-q_2\pd{}{q_1}+p_1\pd{}{p_2}-p_2\pd{}{p_1}\,.
$$
Since 
$$
i(X)\om_\vartheta
=p_2\,dq_1-p_1\, dq_2-q_2\, dp_1+q_1\, dp_2+\vartheta(p_1\,dp_1+p_2\,dp_2)\,,
$$
we find that the generating function is given by 
$$
f(q_1,q_2,p_1,p_2)=
q_1\,p_2-q_2p_1+\frac{\vartheta}{2} \left(p_1^2+p_2^2\right)\,.
$$

We now apply Souriau's minimal coupling procedure, so we introduce 
a basic 2-form $\sg=B(q_1,q_2)\, dq_1\wedge dq_2$, which is closed and can 
be interpreted as a magnetic field, and consider the closed 2-form 
$$
\om_{\vartheta,\sg}:=\om_{\vartheta}-\pi^*\sg\,.
$$
The corresponding linear map
$\om^\flat_{\vartheta,\sg}\:A^1(T^*\R^2)\to\Om^1(T^*\R^2)$ is represented 
by the matrix 
$$
\om^\flat_{\vartheta,\sg}
=\begin{pmatrix}
\hskip 3mm 0&B&-1&\hskip 3mm 0\\ 
-B&0&\hskip 3mm 0&-1\\ 
\hskip 3mm 1&0&\hskip 3mm 0&\hskip 3mm \vartheta\\
\hskip 3mm 0&1&-\vartheta&\hskip 3mm 0
\end{pmatrix},
$$
whose determinant is $(1-\vartheta\, B)^2$, therefore  
$\om_{\vartheta,\sg}$ 
is regular in the points where  $B\,\vartheta\ne1$;
the inverse matrix being  given by 
$$
\om^\sharp_{\vartheta,\sg}=\frac1{1-\vartheta\,B}
\begin{pmatrix}
\hskip 3mm 0&\hskip 3mm \vartheta&\hskip 3mm 1&0\\
- \vartheta&\hskip 3mm 0&\hskip 3mm 0&1\\
-1&\hskip 3mm 0&\hskip 3mm 0&B\\
\hskip 3mm 0&-1&-B&0
\end{pmatrix},
$$
which corresponds to the Poisson structure defined by the bi-vector
$$
\La_{\vartheta,\sg}=\frac1{1-\vartheta\,B}\left(\vartheta \pd{}{q_1}\w
\pd{}{q_2}+\pd{}{q_1}\w \pd{}{p_1}
+\pd{}{q_2}\w \pd{}{p_2}+B\pd{}{p_1}\w \pd{}{p_2}\right).
$$
The corresponding fundamental Poisson brackets are
\begin{align}
\{q_1,q_2\}&=\frac{\vartheta}{1-\vartheta\,B}\,,
&\{q_1,p_1\}&=\{q_2,p_2\}=\frac{1}{1-\vartheta\,B}\,,\notag\\
\{p_1,p_2\}&=\frac{B}{1-\vartheta\,B}\,,
&\{q_1,p_2\}&=\{q_2,p_1\}=0 \,.
\label{eq:previousbrackets}
\end{align}
When the Hamiltonian is  
$$
H=\frac{\mathbf{p}^2}{2m}+V(q_1,q_2)\,,
$$ 
using~\eqref{eq:localformula},~\eqref{eq:timeevolution} 
and~\eqref{eq:previousbrackets}, we see that the time evolution is given by 
\begin{align*}
\dot q_1&=\frac{\vartheta}{1-\vartheta\,B}\,\pd{V}{q_2}
+\frac{p_1}{m(1-\vartheta\,B)}\,,
&\dot p_1&=-\frac{1}{1-\vartheta\,B}\,\pd{V}{q_1}
+\,\frac{Bp_2}{m(1-\vartheta\,B)}\,,\\
\dot q_2&=-\frac{\vartheta}{1-\vartheta\,B}\,\pd{V}{q_1}
+\frac{p_2}{m(1-\vartheta\,B)}\,,
&\dot p_2&=-\frac{1}{1-\vartheta\,B}\,\pd{V}{q_2}
-\,\frac{Bp_1}{m(1-\vartheta\,B)}\,.
\end{align*}
The system is still invariant under rotations if $B$ is a rotationally
invariant function, i.e. $B$ is a function of $q^2_1+q^2_2$, $B=B(q^2_1+q^2_2)$, 
and the generating function for the infinitesimal generator of rotations is 
$$
f(q_1,q_2,p_1,p_2)
=q_1\,p_2-q_2p_1+\frac\vartheta2 (p_1^2+p_2^2)+\frac 12\,B(q_1^2+q_2^2).
$$
On the other hand, when  $B$ is constant, and $B\,\vartheta=1$, the 
determinant of $\omega^\flat_{\vartheta,\sg}$ is zero and the rank of  
$\omega^\flat_{\vartheta,\sg}$ is two, the kernel of the 2-form
$\om_{\vartheta,\sg}$ being generated by the vector fields 
$$
X_1=\vartheta \pd{}{q_2}+\pd{}{p_1}\,,
\sepword{and} X_2=-\vartheta \pd{}{q_1}+\pd{}{p_2}\,.
$$ 
The solutions of $X_1F=X_2F=0$ are to be found from the method of 
characteristics and turn out to be the functions which depend on
$\xi_1=q_1+\vartheta \,p_2$ and $\xi_2=q_2-\vartheta \,p_1$. This suggests 
the change of variables $(q_1,q_2,p_1,p_2)\mapsto (\xi_1,\xi_2,p_1,p_2)$,
i.e. $q_1=\xi_1-\vartheta\, p_2$, $q_2=\xi_2+\vartheta\, p_1$. In such 
coordinates, $X_1=\partial/\partial p_1$, $X_2=\partial/\partial p_2$
and $\om_{\vartheta,\sg}$ becomes
$$
\om_{\vartheta,\sg}=-B\,d\xi_1\w d\xi_2\,.
$$
This means that the quotient manifold $T^*\R^2/\ker\om_{\vartheta,\sg}$ 
is parametrized by $\xi_1$ and $\xi_2$ which, moreover, are Darboux 
coordinates for such 2-dimensional symplectic manifold. 

Using the same idea with commutators Nair and 
Polychronakos~\cite{NairPolychronakos} described quantum mechanics for both 
the non-commutative plane and the non-commutative sphere, and proved that 
the Landau problem for the non-commutative plane can be recovered as the 
limit of large radius of the Landau problem for the non-commutative sphere.

\subsection{Nonrelativistic anyon model in Souriau formalism}
                                                                                     
Let $(q_1,q_2)$ be orthogonal Cartesian coordinates in $Q={\R}^2$, and 
consider the Lagrangian $L_0$ in $\F(TQ)$ of the free particle  
$$
L_0(q_1,q_2,v_1,v_2)=\frac 12\, m\bigl(v_1^2+v_2^2\bigr)\,.
$$
Let $M$ be the graph of the corresponding Legendre transformation.  This is 
the submanifold of the Pontryagin bundle $TQ\oplus T^*Q$ given by the 
constraint functions
$$
\lambda_i(q_1,q_2,v_1,v_2,p_1,p_2)= p_i-m\, v_i\,.
$$
Let  $\kappa\in\mathbb{R}$ be a constant and consider  $TQ$ endowed with the 
 exact 2-form $\omega_1$ defined by
$$
\omega_1(q_1,q_2,v_1,v_2):=\kappa\, dv_1\wedge dv_2\,.
$$
In the spirit of Souriau's idea,  we consider the closed 2-form on 
$TQ\oplus T^*Q$
$$
\om:=- {\rm pr}_1^* \omega_1+{\rm pr}_2^* \omega_0\,,
$$
where ${\rm pr}_1$ and ${\rm  pr}_2$ are the natural projections 
${\rm pr}_1:TQ\oplus T^*Q\to TQ$ and ${\rm pr}_2:TQ\oplus T^*Q\to T^*Q$,
$\om_0$ is the canonical symplectic structure in $T^*Q$ and
$\om_1\in\Om^2(TQ)$ is as before. The corresponding map 
$\om^\flat:A^1(TQ\oplus T^*Q)\to\Om^1(TQ\oplus T^*Q)$ is represented by
the matrix 
$$
\om^\flat=\begin{pmatrix} 
0&0&\hskip 3mm 0& 0&-1&\hskip 3mm 0\\ 
0&0&\hskip 3mm 0&0&\hskip 3mm 0&-1\\ 
0&0&\hskip 3mm 0&\kappa&\hskip 3mm 0&\hskip 3mm 0\\
0&0&-\kappa&0&\hskip 3mm 0&\hskip 3mm 0\\
1&0&\hskip 3mm 0&0&\hskip 3mm 0&\hskip 3mm 0\\
0&1&\hskip 3mm 0&0&\hskip 3mm 0&\hskip 3mm 0
\end{pmatrix},
$$
which is regular for $\kappa\ne 0$. In this case, the 2-form $\om$ is
symplectic, and since the inverse matrix is
$$
\om^\sharp=\begin{pmatrix} 
\hskip 3mm 0&\hskip 3mm 0&0&\hskip 3mm 0&1&0\\ 
\hskip 3mm 0&\hskip 3mm 0&0&\hskip 3mm 0&0&1\\ 
\hskip 3mm 0&\hskip 3mm 0&0&-1/\kappa&0&0\\
\hskip 3mm 0&\hskip 3mm 0&1/\kappa&\hskip 3mm 0&0&0\\
-1&\hskip 3mm 0&0&\hskip 3mm 0&0&0\\
\hskip 3mm 0&-1&0&\hskip 3mm 0&0&0
\end{pmatrix},
$$
we obtain the bi-vector field on $TQ\oplus T^*Q$
$$
\La=\pd{}{q_1}\wedge \pd{}{p_1}+\pd{}{q_2}\wedge \pd{}{p_2}
-\frac1\kappa \pd{}{v_1}\w\pd{}{v_2}\,,
$$
and the corresponding Poisson structure is defined by the following 
fundamental relations:
\begin{align*}
\{q_1,p_1\}&=\{q_2,p_2\}=1,
&\{q_1,q_2\}&=\{p_1,p_2\}=\{q_1,p_2\}=\{p_1,q_2\}=0\,,  \\
\{v_1,v_2\}&=-\frac 1\kappa\,, &\{v_1,p_2\}&=\{v_2,p_1\}=0\,.
\end{align*}
The two constraint functions $\lambda_1$ and $\lambda_2$ are second class
constraints, because
$$
\{\lambda_1,\lambda_2\}=\{p_1-mv_1,p_2-mv_2
\}=m^2\,\{v_1,v_2\}=-\frac{m^2}{\kappa}\,.
$$
They define a four dimensional symplectic manifold.

\section{ Feynman--Dyson's method and non-commutativity}

In this section we review the Feynman's derivation of Maxwell's 
equations~\cite{Dyson}, in the framework of a tangent bundle, so the
Poisson manifold $M$ is the tangent bundle $TQ$ of a configuration
space $Q$. In terms of local tangent bundle coordinates in $TQ$ 
induced from local coordinates in $Q$, denoted  $x_i$ and $\dot x_i$,
a general Poisson bracket on $TQ$ is locally given by
\begin{equation}
\{f,g\}=\{x_a,x_b\}\pd{g}{x_b}\pd{f}{x_a}
+\{x_a,\dot x_b\}\pd{g}{\dot x_b}\pd{f}{x_a}
+\{\dot x_a,x_b\}\pd{g}{x_b}\pd{f}{\dot x_a}
+\{\dot x_a,\dot x_b\}\pd{g}{\dot x_b}\pd{f}{\dot x_a}\,.
\label{eq:Genbracket}
\end{equation}
The asumptions in~\cite{Dyson} are Newton's equations of motion
$$
m\ddot x_j = F_j(x,\dot x),
$$
i.e. the dynamics is given by the second order differential equation 
vector field
$$
\Gamma= \dot x_i\pd{}{x_i}+ F_i(x,\dot x)\pd{}{\dot x_i},
$$
together with the fundamental brackets
\begin{equation}
\{x_i,x_j\} = 0 \sepword{and}
m\{x_i,\dot x_j\} = \dl_{ij}. 
\label{eq:Fbrackets}
\end{equation}
The goal is to determine the other fundamental Poisson brackets, and as 
$\{\dot x_i,\dot x_j\}$ must be skew symmetric it can be written as
$$
\{\dot x_i,\dot x_j\}=  \frac{1}{m^2} \eps_{ijk}B_k(x,\dot x),
$$
where $\eps_{ijk}$ denotes the fully skew-symmetric Levi--Civita 
tensor and $\bf B$ is defined as the magnetic field.
Now (\ref{eq:Fbrackets}) implies that
$$
\{x_i,F_j\} = \frac 1m\pd{F_j}{\dot x_i},
$$
and using the derivation property for the time derivative of the second 
equation in~\eqref{eq:Fbrackets}, i.e. assuming that  the vector field 
$\Gamma$ is a derivation of the Poisson structure, we get
\begin{equation}
\{\dot x_i,\dot x_j\} = -\frac{1}{m}\{x_i,F_j\}
=  \frac{1}{m^2} \eps_{ijk}B_k(x,\dot x)\,,
\label{eq:FandB}
\end{equation}
and the Jacobi identity for the functions $x_i,\dot x_j, \dot x_k$, 
$$
\{x_i,\{\dot x_j,\dot x_k\}\}+\{\dot x_k,\{x_i,\dot x_j\}\}
+\{\dot x_j,\{\dot x_k,x_i\}\}=0
$$
entails
$$
0=\{x_r,B_s\} = \frac{1}{m}\pd{B_s}{\dot x_r}.
$$
In particular $\bf B$ does not depend on the dotted variables.
Moreover from the Jacobi identity with three different velocities
we obtain
$$
\Div {\bf B}= 0,
$$
which reveals that the flux of the field $\mathbf{B}$ through a closed
surface is zero, and that magnetic monopoles do not exist!

On the other hand, since $\bf B$ does not depend on $\dot x_i$, 
equations~\eqref{eq:Genbracket} and~\eqref{eq:FandB}  entail
that $\bf F$ is at most linear in such variables, therefore 
we can define another field ${\bf E}$, called the electric field, by 
$E_j = F_j -\eps_{jkl}\dot x_kB_l$. Using repeatedly all the
equations above, one arrive to Maxwell's equation corresponding 
to Faraday's law of electrodynamics, in the setting suggested 
at the beginning of this section, namely $\Rot {\bf E} = 0$
in the autonomous case, or in general 
$$
\pd{\bf B}{t}+\Rot {\bf E} = 0,
$$
a magnetic field that is changing in time produces a non-conservative
electric field.
We refer the reader to~\cite{Bracken,Bracken2} for details.

To obtain a dynamic different from the standard formalism of
electrodynamics, one needs to modify the fundamental brackets 
in~\eqref{eq:Fbrackets}. The first idea is to use Souriau's
technique, namely to replace the first fundamental bracket by
$$
\{x_i,x_j\} = g_{ij}(x),
$$
where $g_{ij}$ is an arbitrary skew-symmetric matrix of functions, 
fulfilling the constraints that the Poisson bracket properties impose,
and keep the other assumptions. In particular, the Jacobi identity
$$
\{x_i,\{x_j,\dot x_k\}\}+\{\dot x_k,\{x_i,x_j\}\}
+\{x_j,\{\dot x_k,x_i\}\}=0,
$$
entails 
\begin{equation}
0=\{\dot x_k,g_{ij}\}=\{\dot x_k,x_l\}\pd{g_{ij}}{x_l}
+ \{\dot x_k,\dot x_l\}\pd{g_{ij}}{\dot x_l}
=-\frac{1}{m}\pd{g_{ij}}{x_k}.
\label{eq:gwrtx}
\end{equation}
Then  the matrix $g_{ij}$ is a constant skew-symmetric $3\times 3$ matrix,
which is an interesting, but somewhat restrictive, case. We then
modify Souriau's idea and settle for
$$
\{x_i,x_j\} = g_{ij}(x,\dot x).
$$
Accordingly,~\eqref{eq:gwrtx} becomes 
$$
0=\{\dot x_k,g_{ij}\}=-\frac{1}{m}\pd{g_{ij}}{x_k}
+ \{\dot x_k,\dot x_l\}\pd{g_{ij}}{\dot x_l}.
$$
This equation clearly relates the part of the Poisson structure
on the base (the positions) with the part of the Poisson structure
on the fibre (the velocities). Hence if the  Poisson structure on 
the base is known one can compute the fundamental brackets on the 
fibre.

On the other hand, from the Jacobi identity among
$(x_i, x_j, x_k)$ we obtain
$$
\{x_i,g_{jk}\} + \{x_k,g_{ij}\} + \{x_j,g_{ki}\} = 0,
$$
and this leads to another constraint:
\begin{equation}
0 =g_{il}\pd{g_{jk}}{x_l} + g_{kl}\pd{g_{ij}}{x_l}
+g_{jl}\pd{g_{ki}}{x_l} +\frac{1}{m}\Bigl(
\pd{g_{jk}}{\dot x_i} +\pd{g_{ij}}{\dot x_k}
+\pd{g_{ki}}{\dot x_j}\Bigr).
\label{eq:funnyeq}
\end{equation}
Note that if $d_x$ denotes the exterior derivative on the
vector space $T_xQ$, for $x$ in $Q$, then the term inside the
parenthesis in~\eqref{eq:funnyeq} are the local coordinates of 
$d_x\tilde\om$, where $\tilde\om_x$ is the 2-form in $\Om^2(T_xQ)$ 
defined by $\tilde\om_x= g_{ij}(x,\dot x)d\dot x_i\w d\dot x_j$. On 
the other hand, if $\tilde\La$ is the bivector in $A^2(TQ)$ given by
$\tilde\La = g_{ij}(x,\dot x)\pd{}{x_i}\w\pd{}{x_j}$ the terms
outside the parenthesis in~\eqref{eq:funnyeq} are the local coordinates
of $[\tilde\La,\tilde\La]_{SN}$. Therefore equation~\eqref{eq:funnyeq} 
is fulfilled when 
$$
d_x\tilde\om=0\sepword{and} 
d_{\tilde\La}\tilde\La=[\tilde\La,\tilde\La]_{SN}=0,
$$
that is, when $\tilde\om_x$ is closed and $\tilde\La$ is a Poisson
tensor.

Furthermore, similar ideas as in the commutative case,
using the other Jacobi identities, leads, see~\cite{Felicia},
to the modified Gauss law 
$$
\Div {\bf B} = -\frac{1}{m}\,{\bf B}\cdot \dot\nabla \x {\bf B},
$$
where $\dot\nabla =(\pd{}{\dot x_1},\pd{}{\dot x_2},\pd{}{\dot x_3})$,
and also to 
$$
({\rm rot}\, {\bf E})_k+\frac 1m\left(({\bf E}\cdot \dot \nabla)B_k
+{\bf B}\cdot \pd{{\bf E}}{\dot x^k}-(\dot\nabla\cdot {\bf E})\,B_k\right)=0\,,
$$
which is what replaces the Maxwell equation corresponding to Faraday's law.

\subsection{ Generalized Lorentz force equations}

Consider the Hamiltonian dynamical system on $T\R^3$, where
the Hamiltonian and the symplectic form are given respectively by
$$
H = \frac{1}{2m}\delta_{ij}\dot{x}_{i}\dot{x}_{j}
+ \phi(x),
$$
and 
$$
\om = \frac{1}{m} dx_i\w d\dot{x}_{i} 
+B_1 dx_2\w dx_3 + B_2 dx_3\w dx_1 + B_3 dx_1\w dx_2
+ \frac{1}{2}g_{ij}d\dot{x}_{i}\w d\dot{x}_{j}.
$$
Assume that, in local coordinates, the Hamiltonian vector field is
written as $X_H = S_i\partial_{x_i}+R_i\partial_{\dot x_i}$. The equation
$i(X_H)\Om=dH$ becomes
\begin{align}
\frac{1}{m}\dot x_1 &=\frac{1}{m}S_1+g_{21}R_2+g_{31}R_3,
&\pd{\phi}{x_1}&=-\frac{1}{m}R_1+B_2S_3-B_3S_2,  
\notag\\
\frac{1}{m}\dot x_2 &=\frac{1}{m}S_2+g_{12}R_1+g_{32}R_3,
&\pd{\phi}{x_2}&=-\frac{1}{m}R_2+B_3S_1-B_1S_3,  
\label{eq:veinticuatro}\\ 
\frac{1}{m}\dot x_3 &=\frac{1}{m}S_3+g_{13}R_1+g_{23}R_2,
&\pd{\phi}{x_3}&=-\frac{1}{m}R_3+B_1S_2-B_2S_1 .
\notag
\end{align}
On the other hand, from~\eqref{eq:timeevolution} 
and~\eqref{eq:Pbracketagain} we obtain
$$
\fd{x_i}{t}=S_i \sepword{and}\fd{\dot x_i}{t}=R_i.
$$
Therefore if we assume that $X_H$ is a second order differential 
equation, namely that $S_i=\dot x_i$, then $\sd{x_i}{t}=R_i$, and
the right column of~\eqref{eq:veinticuatro} entails
\begin{align}
\frac{1}{m}\sd{x_1}{t}&=-\pd{\phi}{x_1} +\dot x_3B_2-\dot x_2B_3\,,
\notag\\
\frac{1}{m}\sd{x_2}{t}&=-\pd{\phi}{x_2} +\dot x_1B_3-\dot x_3B_1\,,
\label{eq:tres}\\ 
\frac{1}{m}\sd{x_3}{t}&=-\pd{\phi}{x_3} +\dot x_2B_1-\dot x_1B_2\,,
\notag
\end{align}
whereas the left column provides the constraints
\begin{align*}
0&=mg_{21}\Bigl(\pd{\phi}{x_2}-\dot x_1B_3+\dot x_2B_1\Bigr)
+ mg_{31}\Bigl(\pd{\phi}{x_3}-\dot x_2B_1+\dot x_1B_2\Big), \\
0&=mg_{12}\Bigl(\pd{\phi}{x_1}-\dot x_3B_2+\dot x_2B_3\Bigr)
+ mg_{32}\Bigl(\pd{\phi}{x_3}-\dot x_2B_1+\dot x_1B_2\Bigr), \\
0&=mg_{13}\Bigl(\pd{\phi}{x_1}-\dot x_3B_2+\dot x_2B_3\Bigr)
+ mg_{23}\Bigl(\pd{\phi}{x_2}-\dot x_1B_3+\dot x_3B_1\Bigr).
\end{align*}
In particular when $\nabla\phi=-eE$, \eqref{eq:tres} is a 
generalized Lorentz force: a force experienced by a charged
particle moving in an electromagnetic field, subject to
a system of constraints.

Another interesting class of systems can be studied via 
the ``generalized Souriau form''
$$
\tilde\om_0= d\dot x_i\w dx_i+g_{ij}d\dot x_i\ox d\dot x_j.
$$
It is a mixture of a symplectic and a gradient structure,
known as a metriplectic system. The symmetric bracket
associated to the metric tensor incorporates the dissipative
structure of the system. The Leibniz vector field $X_h$ associated
to a function $h \in C^{\infty}(M)$ satisfies $X_h = \nabla h$,
i.e. $X_h$ generates a gradient dynamical system. In local coordinates
the vector field $X_h$ is given by
$$
X_h=g_{ij}\pd{h}{x_j}\pd{}{x_i}\,,
$$
and the corresponding bracket in this context is called a Leibniz bracket.

\section{Volume preserving mechanical system related to Souriau form}

Another interesting class of dynamical systems that generalize the
Hamiltonian systems, where noncommutativity is also possible, was introduced 
in~\cite{ZhouGuoWu,ZhouGuoPanWu}.
 
Let $(M,\om_0)$ be a $2n$-dimensional symplectic manifold, a vector field
$X$ is said to be \textit{symplectic} if $\mathcal{L}_X\om_0=0$, from the Cartan
identity, this is equivalent to $i(X)\om_0$ being closed, in particular every
Hamiltonial vector field is symplectic. On the other hand, we say a vector
field $X$ \textit{preserves the volume} if $\mathcal{L}_X\om_0^n=0$; here and in
what follows powers are meant with respect to the wedge product. Since
$$
\fd{}{s}\Phi^*_{t+s}\om^n_0\Big|_{s=0}
=\Phi_t^*\fd{}{s}\Phi^*_s\om^n_0\Big|_{s=0}
=\Phi_t^*\mathcal{L}_X\om^n_0=0,
$$
where $\Phi_t$ is the flow of $X$, it follows that $\Phi_t$ does preserves
the volume form $\om_0^n$. Furthermore, a simple induction gives
$$
\mathcal{L}_X\om^k_0=k\mathcal{L}_X\om_0\w\om^{k-1}_0.
$$
In particular, we see that every symplectic vector field preserves the
volume, but the converse is not true in general. The divergence of a vector
field $X$ is defined as the unique function $\Div X$ in $C^\infty(M)$ such
that 
$$
\mathcal{L}_X\om^n_0=\Div X\,\om^n_0. 
$$
Therefore $X$ preserve the volume if, and only if, it is divergence free.
Let $(x_1,\dots,x_{2n})$ be Darboux coordinates, then 
$\om_0= dx_i\w dx_{n+i}$, 
and if $X=\sum_{i=1}^{2n}X_i\partial_{x_i}$ it is easy to check that
$$
\Div X=\sum_{i=1}^{2n}\pd{X_i}{x_i}\,.
$$

We now describe a procedure that produces dynamical systems that
preserves the volume. First consider the map $F\:A^1(M)\to\Om^{2n-1}(M)$
given by $F(X):=i(X)\om^n_0$. Using Darboux coordinates, simple 
combinatorial arguments entail 
\begin{align}
\om^k=(-1)^{\frac{k(k-1)}{2}}k!\sum_{1\le i_1<\cdots<i_{n-k}\le n}\Bigl(
dx_1\w dx_{n+1}&\w\dots\w  \widehat{dx_{i_1}}\w\widehat{dx_{n+i_1}}\w\dots
\notag\\
&\w\widehat{dx_{i_{n-k}}}\w\widehat{dx_{n+i_{n-k}}}\w\dots \w dx_n\w dx_{2n}
\Bigr),
\label{eq:powers}
\end{align}
where as usual the hat means that the term is to be deleted, in particular
\begin{align}
i(X)\om^n_0&=(-1)^{\frac{n(n-1)}{2}}n!i(X)dx_1\wyw dx_{2n}
\notag\\
&=n!(-1)^{\frac{n(n-1)}{2}+i-1}X_i\,dx_1\wyw\widehat{dx_i}\wyw dx_{2n},
\label{eq:contraction}
\end{align}
therefore $F$ is surjective. Moreover, if $X(p)\ne0$ and $F(X)=0$, we can 
locally find a basis $X_1,\dots,X_{2n}$, of vector fields such that $X_1=X$, 
then
$$
\om_0^n(X_1,\dots,X_{2n})=i(X)\om_0^n(X_2,\dots,X_{2n})=0,
$$
which is absurd since $\om_0$ is nondegenerate, hence $F$ is injective
and therefore a linear isomorphism. Since $\mathcal{L}_X\om^n_0=di(X)\om^n_0$, 
under $F$, the space of volume preserving vector fields corresponds to 
the space of closed $(2n-1)$-forms. Thus, if $\eta$ is a 2-form 
$$
X_\eta:=F^{-1}\bigl(d(\eta\w\om_0^{n-2})\bigr)
$$ 
is a volume preserving vector field. In particular, if $\om$ is the
Souriau form, written in Darboux coordinates as
$$
\om= dx_{n+i}\w dx_i+\frac{1}{2}g_{ij}x_{n+i}\w dx_{n+j}
+\frac{1}{2}B_{ij} dx_i \w dx_j,
$$
then 
\begin{align*}
d(\om\w\om_0^{n-2})=d\om\w\om_0^{n-2}
&=\frac{1}{2}\,\pd{g_{ij}}{x_k}\,dx_k\w dx_{n+i}\w dx_{n+j}\w\om_0^{n-2} \\
&\quad+
\frac{1}{2}\,\pd{B_{ij}}{x_{n+k}}\,dx_{n+k}\w dx_i\w dx_j\w\om_0^{n-2}.
\end{align*} 
Now, from  \eqref{eq:powers}
\begin{align*}
 dx_k\w dx_{n+i}\w dx_{n+j}\w\om_0^{n-2}&=
\dl_{ki}\;dx_{n+j}\w dx_k\w dx_{n+i}\w\om_0^{n-2}  \\
&\quad-\dl_{kj}\;dx_{n+i}\w dx_k\w dx_{n+j}\w\om_0^{n-2}  \\
&=\frac{(-1)^n}{n-1}\Bigl( \dl_{ki}\; dx_{n+j}\w\om_0^{n-1}
-  \dl_{kj}\;dx_{n+i}\w\om_0^{n-1}\Bigr).
\end{align*}
Similarly
$$
 dx_{n+k}\w dx_i\w dx_j\w\om_0^{n-2}
 =\frac{(-1)^n}{n-1}\Bigl( \dl_{kj}\;dx_i\w\om_0^{n-1}
  -  \dl_{ki}\;dx_j\w\om_0^{n-1}\Bigr).
$$
Thus
$$
d(\om\w\om_0^{n-2})=\frac{(-1)^n}{n-1}\left(
\pd{g_{ij}}{x_i}\,dx_{n+j}\w\om_0^{n-1}
-\pd{B_{ij}}{x_{n+k}}\,dx_j\w\om_0^{n-1}\right).
$$
Let $X$ be the vector field
$$
X=c_n\;\pd{g_{kl}}{x_k}\;\pd{}{x_l}
-c_n\;\pd{B_{kl}}{x_{n+k}}\;\pd{}{x_{n+l}}, \sepword{with} 
c_n=\frac{(-1)^{\frac{n(n-1)}{2}}}{n(n-1)},
$$
then, using \eqref{eq:contraction} and \eqref{eq:powers}
\begin{align*}
i(X)\om_0^n&=n!(-1)^{\frac{n(n-1)}{2}+i-1}c_n\;\pd{g_{ki}}{x_k}\;
dx_1\w\cdots\w\widehat{dx_i}\w\cdots\w dx_{2n} \\
&\quad-n!(-1)^{\frac{n(n-1)}{2}+n+i-1}c_n\;\pd{B_{ki}}{x_{n+k}}\;
dx_1\w\cdots\w\widehat{dx_{n+i}}\w\cdots\w dx_{2n} \\
&=(n-2)!(-1)^n\;\pd{g_{ki}}{x_k}\;dx_{n+i}\w
dx_1\w\cdots\w\widehat{dx_i}
\w\cdots\w\widehat{dx_{n+i}}\w\cdots\w dx_{2n} \\
&\quad-(n-2)!(-1)^n\;\pd{B_{ki}}{x_{n+k}}\;dx_i\w
dx_1\w\cdots\w\widehat{dx_i}
\w\cdots\w\widehat{dx_{n+i}}\w\cdots\w dx_{2n} \\
&=(n-2)!(-1)^{n+\frac{(n-1)(n-2)}{2}}\left(\pd{g_{ki}}{x_k}\;dx_{n+i}
-\pd{B_{ki}}{x_{n+k}}\;dx_i\right) \\
&\hskip 5cm\w
dx_1\w dx_{n+1}\w\cdots\w\widehat{dx_i}\w\widehat{dx_{n+i}}
\w\cdots\w dx_n\w dx_{2n} \\
&=d(\om\w\om_o^{n-2}).
\end{align*}
In other words, $X_\om=X$, so we can associate a volume preserving
flow with the Souriau's form, and the equations of motion are given by
\begin{align*}
\fd{x_i}{t} &= \frac{(-1)^n}{n-1}\;\pd{ g_{ki}}{x_k}\,,  \\
\fd{x_{n+i}}{t} &=-\;\frac{(-1)^n}{n-1}\;\pd{B_{ki}}{x_{n+k}}\,.
\end{align*}

A Nambu-Poisson system is a volume preserving flow, 
on a Nambu-Poisson manifold of order $2n$, determined by $(2n-1)$ 
Hamiltonian functions $H_1,\dots,H_{2n-1} \in C^{\infty}(M)$
$$
\fd{x_i}{t}= X_{H_1,\dots,H_{2n-1}}(x_i) = \{H_1,\dots,H_{2n-1}, x_i \}.
$$
In general, $\mathcal{L}_{X_{H_1,\dots,H_{2n-1}}}f = \{H_1,\dots,H_{2n-1},f \}$ if 
$X_{H_1,\dots,H_{2n-1}}$ is a Nambu-Hamiltonian vector field. 
If $\eta$ is a Nambu-Poisson 
tensor, then Takhtajan \cite{Takhtajan} proved 
$\mathcal{L}_{X_{H_1,\dots,H_{2n-1}}}\eta = 0$.

\section{Conclusion and Outlook}

In this paper we have studied the classical non-commutative 
mechanical systems using Souriau's method of generalized symplectic form.
In particular we have explored a large class of non-commutative flows which 
includes the non-commutative magnetic geodesic flows, non-relativistic anyon 
model \cite{HP02,HP04,HP05}, generalized Lorentz force equation, etc. Souriau's formalism allows
us to study geometrically all these non-commutative dynamical systems
in an unified manner. The dynamics of these systems boil
down to generalized Hamiltonian dynamics where the Poisson structure
can be complicated functions of phase space coordinates and momenta.
However, some questions should be addressed in near future.

At first we must consider the quantization of these classical non-commutative
system. There is an interesting paper \cite{BMLGGBM} which addresses the 
connection between non-commutative quantum mechanics and Feynman--Dyson's method.
Actually, the generalization of Feynman--Dyson's idea to the quantum world
would be an interesting subject to be studied. This would lead to unveil
the close relation existing between the non-commutative geometry
and the geometric phases. Quantization of these models could give
rise to new physics at some very high energy scale \cite{{BBDP14},{DL06},{DL08},{FM04},{FSK14},{GSS14a},{GSS14b},{J08},{K12},{Martina3},{PG13}, {S12}}.

This method can be extended to other directions also.
We can simply generalize this construction to
supersymmetric  non-commutative systems \cite{HPV07,HPV10}. In other words, we
can try to generalize the Feynman--Dyson's scheme to
supersymmetric framework.
This would certainly yield a generalization of supersymmetric
generalized Hamiltonian dynamics.
There is a recent upsurge of interest in $(2+1)$-dimensional model \cite{FVG,V14} with a kind of nonstandard noncommutativity,
where both coordinates and momenta get deformed commutators.
So far most of the papers concern time-independent systems, it would be rather 
challenging to extend this framework to time-dependent systems. In a recent note
Liang and Jiang \cite{LJ10} studied the time-dependent harmonic oscillator in a background
of time-dependent electric and magnetic fields. Recently noncommutative quantum mechanics \cite{BMLGGBM,GLR,BGK,AA,{SGHR}}
is becoming a exciting topic to study, it would be interesting for us to investigate
this subject using geometrical methods of quantizing noncommutative phase space mechanics.

\smallskip

Possibly one can study the Helmholtz condition analogous to
standard classical dynamics. Then the corresponding variational
formulation can be used to construct Noether symmetries and conserved densities.
It is known \cite{CorteseGarcia} that the Helmholtz condition connected to 
$\theta$-deformed Poisson system is a third-order time derivative equation. 
One should analyse carefully all these new aspects.

We can also study the field theoretic Poisson brackets
on jet space. This will yield an interesting class
of partial differential equations. One must try to explore
its connection to other branches of mechanics and geometry, namely,
non-holonomic systems, control theory, Finslerian mechanics,
Lie algebroid theory, etc. 

\subsection*{Acknowledgments}

We thank (Late) Jerry Marsden, Christian Duval, Peter Horv\'athy,
Parameswaran Nair, Ali Chamse\-ddine and Frederik Scholtz for useful 
conversations. JFC acknowledges financial support from research projects 
MTM--2012--33575 (MINECO, Madrid) and DGA-GRUPOS CONSOLIDADOS E/21.
HF acknowledges support from the Vicerrector\'{\i}a de 
Investigaci\'on of the Universidad de~Costa~Rica. HF and PG
thank the Departamento de F\'{\i}sica Te\'orica de la 
Universidad de Zaragoza for its warm hospitality.
The final part of the work was done while PG was visiting IHES. He would 
like to express his gratitude to the members of IHES for their warm hospitality.

\end{document}